\documentclass[english,preprint,12pt]{aastex}
\usepackage[LGR,T1]{fontenc}
\usepackage{color}
\usepackage{amstext}
\usepackage{graphicx}

\makeatletter

\slugcomment{}
\shorttitle{Scaling of properties in rotating stars}
\shortauthors{Castaneda D. et al.}

\makeatother

\usepackage{babel}
\begin{document}

\title{Scaling of Observable Properties in Rotating Stars}

\author{D. Castañeda, R. G. Deupree}

\affil{Institute for Computational Astrophysics and Department of Astronomy
and Physics}

\affil{St. Mary's University, Halifax, NS B3H 3C3, Canada}

\email{castaned@ap.smu.ca}
\begin{abstract}
The spectral energy distribution as a function of inclination is computed
using 2D rotating stellar models and NLTE plane parallel stellar atmospheres.
These models cover the range from $1.875M_{\odot}$ to $3.0M_{\odot}$.
The deduced effective temperature is determined by B-V computed from
the spectral energy distribution, and the deduced luminosity is computed
as the integral of the spectral energy distribution over all frequencies,
assuming the distance and reddening are known. These deduced quantities
are obtained from the observed spectral energy distribution assuming
the objects are spherically symmetric, and thus the results are dependent
on the inclination. Previous work has shown that the surface properties
between two rotating stellar models with the same surface shape scale,
and this is also true for the deduced effective temperature and luminosity
over this limited mass range.
\end{abstract}


\section{INTRODUCTION}

Advances in the general study of rotating stars have been limited
by both theoretical and observational difficulties. A key example
is that fundamental properties such as the effective temperature ($T_{\mathrm{eff}}$)
and luminosity ($L$) that one would deduce from observations now
depend significantly on the angle of inclination ($i$) between the
line of sight and the star\textquoteright{}s rotation axis for sufficiently
rapidly rotating stars (e.g., \citealt{Collins1966,Hardorp1968,Maeder1970}).
This greatly complicates the determination of the star\textquoteright{}s
position on the HR diagram and hence nearly all other useful information
unless its inclination can be determined. 

Given a grid of plane parallel model atmospheres and a rotating model
with latitudinal variation in the surface effective temperature and
effective gravity, one can compute the spectral energy distributions
(SED) for any inclination by performing a weighted integral of the
intensity in the direction of the observer over the visible surface
of the star (e.g., \citealt{Slettebak1980,Linnell1994,Fremat2005a,Lovekin2006a}).
Integration of the flux in the SED over wavelength produces what we
will refer as deduced $L$, and the application of nonrotating color
\textendash{} effective relations to the SED produces what we will
denote as deduced $T_{\mathrm{eff}}$ Both these quantities will be
strongly inclination dependent for sufficiently rapid rotation. There
is also no straightforward relationship with the actual $L$ (i.e.
the total amount of energy coming out of the star per unit time) or
the actual $T_{\mathrm{eff}}$, which we define as $(L/\left(A\sigma\right))^{1/4}$
where $A$ is the surface area of the star and $\sigma$ is the Stefan\textendash{}Boltzmann
constant. Naturally the comparison between a computed SED and one
observed has much prospect of success only if the inclination and
the oblateness of the star are known. Fortunately important advances
in interferometric instrumentation over approximately the last decade
have permitted resolved observations of some nearby rapid rotators
(e.g., \citealt{VanBelle2001,DomicianodeSouza2003,Aufdenberg2006,Monnier2007,Zhao2009,Che2011}).
The direct process requires performing the calculations with different
models until the observed SED properties are matched to the extent
possible. This can be laborious and it would be far preferable to
be able to start with the observed SED properties and work backward
to what the luminosity and the latitudinal variation of the effective
temperature must be. A crucial first step in this process is to demonstrate
that there is a well defined relation between the deduced quantities
and their physically significant counterparts, even though this relation
depends on both the inclination and the amount of rotation. We explore
this relationship in this work.

Recently \citet{Deupree2011a} showed that a number of properties
of rotating models, particularly the surface effective temperature
as a function of latitude, is proportional between models as long
as the surface shape remains the same. Of course the surface radius
as a function of latitude scales by definition. For the surface shapes
to be exactly the same, the two models must have the same rotation
law to within a multiplicative constant. The independence of latitude
for the actual effective temperature and radius ratios suggests that
observable properties such as the deduced luminosity and deduced effective
temperature as functions of inclination may scale as well. If true,
one might be able to at least place constraints on models and parameters
which could produce the observed properties. It would also allow stepping
backwards from the observed properties to the actual luminosity and
effective temperature in a straightforward way for cases in which
both the inclination and surface shape are known. Being able to reduce
the uncertainty in a star's actual properties is important in determining
whether differences between observed and computed oscillation frequencies
are due to not having a model with the correct actual properties or
not having the interior model structure right. This is very timely
because the discovery of multiple oscillation frequencies in rapidly
rotating stars such as $\alpha$ Oph (\citealt{Monnier2010a}) for
which interferometric observations (\citealt{Zhao2009}) have been
made provides our best opportunity at the moment to study in detail
the interior structure of rotating stars (e.g. \citealt{Deupree2012,IAU:9177147}).

We develop the model scaling relationships and discuss how the deduced
properties scale in the next section. The following sections provide
an example of how these scaling properties can be used and what are
the limitations of the method.

\section{SCALING PROPERTIES}

We wish to examine the surface relationships between models which
have the same surface shape. We assume that the surface is an equipotential,
requiring the rotation law to be conservative. Some of the assumptions
we make in the analysis hold for the laws we consider, and it is not
clear what deviations from a conservative law are allowed before none
of the scaling results hold. We shall return to this point in the
final section.

\subsection{Scaling of Model Properties\label{sub:Scaling-of-Model-properties}}

We first consider what scaling in this context means. Essentially,
it means that for any surface variable of two models ($y_{1}$, $y_{2}$)
with the same shape:\textbf{
\begin{equation}
\frac{y_{2}(\theta_{j})}{y_{1}(\theta_{j})}=c_{y}\quad\forall\quad j\label{eq:1}
\end{equation}
}where $\theta_{j}$ is the co-latitude and $c_{y}$ is a constant.\textbf{
}This is true for the surface radius by definition.

Having the surface shape of two different models the same also imposes
a number of conditions on those models. First, models that have the
same surface shape have the same rotation law form, except for an
overall scale factor:\textbf{
\begin{equation}
\Omega_{2}(X_{R},\theta_{j})=c_{\Omega}\times\Omega_{1}(X_{R},\theta_{j})\quad\forall\quad j
\end{equation}
}where $\Omega$ is the rotation rate, $c_{\Omega}$ is a constant
and\textbf{ $X_{R}=r_{R_{1}}/R_{1}=r_{R_{2}}/R_{2}$.}

The scaling of the rotation rate and the surface radius between two
models with the same surface shape means the rotational velocities
will scale as well. The scale factor can be determined from the ratio
of the surface equatorial velocities of the two models. With both
the rotational velocity and radius scaling, the centrifugal force
scales as well.

Letting $\Phi$ denote the gravitational potential, we can write the
total equipotential ($\Psi$) at the surface in the following way:
\begin{eqnarray}
\Psi_{1} & = & \Phi_{1}(\theta)+\Omega_{1}^{2}(\theta)R_{1}^{2}(\theta)\frac{\sin^{2}(\theta)}{2}=\frac{\Psi_{2}}{c_{\Psi}}=\frac{1}{c_{\Psi}}\left[\Phi_{2}(\theta)+\Omega_{2}^{2}(\theta)R_{2}^{2}(\theta)\frac{\sin^{2}(\theta)}{2}\right]\nonumber \\
 & = & \frac{1}{c_{\Psi}}\left[\Phi_{2}(\theta)+c_{\Omega}^{2}c_{R}^{2}\Omega_{1}^{2}(\theta)R_{1}^{2}(\theta)\frac{\sin^{2}(\theta)}{2}\right]
\end{eqnarray}
Given that the gravitational\textbf{ potential }will have only a small
nonradial component, we see that this equation can be solved only
if both terms are true individually. Thus,
\begin{equation}
\Phi_{2}=c_{\Psi}\Phi_{1}\:\mathrm{and\:}c_{\Psi}=c_{\Omega}^{2}c_{R}^{2}
\end{equation}
We see that the gravitational potentials scale as well. To the extent
one can approximate the gravity at the surface by the Roche potential
(which one can do quite well for all but the most rapid uniformly
rotating models), one can obtain an estimate for the mass once we
know the appropriate scaling constants. 

\citet{Deupree2011a} has shown that scaling as defined in equation
(\ref{eq:1}) is true for the effective temperature for ZAMS models
from $1.875$ to $8M_{\odot}$ with twenty different shapes. These
effective temperature ratios appear to show a maximum variation of
$0.5\%$ over all latitudes. A part of the variation could be due
to the fact that radii are discretized in the 2D finite difference
mesh - a zone is either inside the model or it is not. The surface
at each latitude is taken to be the outer radial boundary of the zone
which the equipotential that describes the surface passes through.

We can see why the the effective temperatures might scale by application
of von Zeipel\textquoteright{}s (1924) law to two rotating homologous
models which have the same shape. This means that the equipotential
surfaces will have the same shape at the same fractional radius. Because
the flux, assumed to be radiative, is perpendicular to the equipotential
surface, one has
\begin{equation}
\left[\frac{F_{\theta}(X_{R},\theta_{j})}{F_{r}(X_{R},\theta_{j})}\right]_{1}=\left[\frac{F_{\theta}(X_{R},\theta_{j})}{F_{r}(X_{R},\theta_{j})}\right]_{2}
\end{equation}
where the subscripts 1 and 2 refer to the two models. This means that
the same fraction of the flux is being diverted from the radial direction
for each of the two models. If this is true at all locations inside
the model, then the distribution of the flux emerging from the model
outer boundary must have the same relative distribution with latitude
in both models. Because the effective temperatures are defined in
terms of the flux emitted from a surface zone, the effective temperatures
of the two models must scale, satisfying equation (\ref{eq:1}).

\textbf{We can also obtain this result from the work of Espinosa Lara
\& Rieutord (2011) , which finds that one may write $T_{\mathrm{eff}}=a\, g_{\mathrm{eff}}^{\beta}$,
where $\beta$ decreases slightly as the model becomes more oblate.
Because $g_{\mathrm{eff}}$ scales, then $T\mathrm{_{eff}}$ scales
as long as one is comparing models that have the same shape as we
are here. We thank the referee for this insight. }

Given sufficient information, one could expect to compute the latitude
independent ratios of the surface radius, effective temperature, surface
rotation velocity, and effective gravity. While these results may
be of some theoretical interest, they would be more beneficial if
properties obtained from observations, i.e. the deduced $T_{\mathrm{eff}}$
and deduced $L$, also scaled.

\subsection{Scaling of observed properties}

Two key variables one wishes to obtain from a star are the actual
$T_{\mathrm{eff}}$ as defined above and $L$. This remains true for
rotating stars, with the complication that neither the temperature
nor the luminosity one would deduce from observations of a rapidly
rotating star directly relate to intrinsic stellar properties because
the deduced properties are strongly inclination dependent (e.g.\citealt{Collins1966,Hardorp1968,Maeder1970,Gillich2008,Dall2011}).
To obtain a deduced effective temperature and luminosity from a spectral
energy distribution for a rotating star requires the same knowledge
about reddening and distance as for a spherical star, so we assume
that this transformation can be performed to some degree and will
address obtaining the deduced luminosity and effective temperature
from dereddened SEDs with a known absolute flux.

The computation of the deduced effective temperature and luminosity
as a function of inclination requires several steps. First, we must
have the surface properties of the model, which here we take from
the suite of ROTORC (\citealt{Deupree1990}, \citeyear{Deupree1995})
ZAMS models computed by \citet{Deupree2011a}. We note that these
models force a relationship between the local effective temperature
and the local surface temperature, unlike von Zeipel's law which assumes
that the surface is an equipotential (and hence constant temperature)
surface while the effective temperature can vary significantly from
pole to equator. The net effect is that the ROTORC models have a flatter
relationship between the effective temperature and effective gravity,
closer to 0.2 instead of the 0.25 of von Zeipel's law. We note that
the behavior and values of our exponent with increasing rotation are
quite similar to those of \citet{2011A&A...533A..43E}. Previous studies
consistently find a lower value preferable (\citealt{Monnier2007,Che2011,Claret2012}).
Second, we must also have the intensities emerging from the surface
for each member of a grid of stellar atmospheres. For the intermediate
mass main sequence models we wish to explore, plane parallel model
atmospheres are satisfactory. The model atmospheres are computed with
the PHOENIX code (\citealt{HauschildtP.H.1999}), and the grid covers
the range in effective temperature from 7500K to 11000K in steps of
250K and in log g from 3.333 to 4.333 in steps of 0.333. The spectrum
was computed from the far ultraviolet to 20000\AA{}, with minimum
wavelength (wavelength interval) maximum wavelength = 600\AA{} (0.005\AA{})
1500\AA{} (0.01\AA{}) 3000\AA{} (0.015\AA{}) 4000\AA{} (0.02\AA{})
6000\AA{} (0.03\AA{}) 8000\AA{} (0.04\AA{}) 12000\AA{} (0.06\AA{})
16000\AA{} (0.08\AA{}) 20000\AA{}. These intervals were chosen to
keep the resolution greater than 250000 below 8000Å and about 200000
above 8000Å. Lines in the four lowest ionization stages of Al, S,
and Fe; in the three lowest of C, N, O, Mg, K, and Ca; in the two
lowest of He, Li, and Na; and in the lowest of H and Ne are computed
in NLTE. More specific details are given by \citet{Gillich2008} and
\citet{Deupree2012}. At lower temperatures than included we would
need to include more species in NLTE, and at higher temperatures photometric
temperature indicators in the visible region of the spectrum become
harder to find. The net result of these two steps is that at any place
on the surface of the rotating model, one can interpolate through
the grid of model atmospheres in log $T_{eff}$ and log $g$ to obtain
the emergent intensity in any direction with respect to the local
vertical. The third step calculates the direction to the observer,
and thus the angle of the observer with respect to the local vertical,
at every point on the surface, and performs the weighted integral
over all the contributions of the intensities from every point on
the surface visible to the observer to obtain the flux the observer
would see. This approach for the third step is rather frequently used
(e.g.,\citealt{Slettebak1980,Linnell1994,Fremat2005a,Gillich2008,Aufdenberg2006,Yoon2008,Dall2011}),
and the specific details in our calculations are outlined by \citet{Lovekin2006a}.
The final SEDs were obtained by using a 50Å wide boxcar filter. Because
of this filtering and the fact that rotation does not affect the equivalent
width, the Doppler shifts were not included in the flux integrals,
making the flux calculation computationally \textquotedblleft{}embarrassingly
parallel\textquotedblright{}. There is an option to include the Doppler
shift when one wishes to compute specific line profiles with no filtering.

SEDs were obtained for uniformly rotating models for five masses ($1.875$,
$2$, $2.25$, $2.5$ and $3M_{\odot}$) and six different rotation
rates characterized by flatness ($1-R_{\mathrm{p}}/R_{\mathrm{eq}}$)
values of 0.112, 0.134, 0.156, 0.180, 0.207 and 0.234. To give an
idea of how much rotation this is in more conventional terms, we note
that the surface equatorial velocities range from about 230 km s$^{-1}$
to about 360 km s$^{-1}$. The most rapidly rotating model was chosen
to keep the minimum effective temperature above 7500K, below which
we would need to include other low ionization potential metals in
NLTE. Models with slower rotation rates were not included because
the pole to equator temperature variation was less than about 1000K.
Some computed effective temperatures for the $1.875M_{\odot}$ models
for the two most oblate calculations were below this lower temperature
limit, and those models were not included. The SEDs were computed
at ten equally spaced inclinations from pole on to equator on.

Both the deduced effective temperature and deduced luminosity were
obtained from the computed SEDs. As expected, we found that B-V provided
a good indicator of the deduced effective temperatures, using a NLTE
PHOENIX model of Vega with the parameters of \citet{Castelli1994}
to calibrate the color indices. The deduced effective temperatures
were obtained from the simulated (B-V) color using the (B-V) - effective
temperature relation for the plane parallel model atmospheres with
the Vega calibration. The gravity used for the (B-V) - effective temperature
relationship for the rotating models was the effective gravity at
the co-latitude which corresponds to the inclination angle. However,
the variation in effective gravity between the equator and pole is
only a little larger than a factor of two, which would lead to a maximum
error in the effective temperature of about $\pm100K$ based on a
comparison of the change in the color \textendash{} effective temperature
relations with gravity for the plane parallel model atmospheres. The
scaling should still be successful because the models at the same
shape have the same effective gravity distribution.

The deduced luminosities were computed by integrating the computed
flux over all wavelengths, including a Rayleigh-Jeans tail from the
end of the calculated wavelengths to infinite wavelength, and multiplying
the result by $4\pi d^{2}$, where $d$ is an
assumed distance to the model from the observer. Because the SED is
inclination dependent, the deduced luminosity and effective temperature
will be also. We also note that, because determining the gravity becomes
part of the scaling algorithm if the inclination and shape are known,
one can iterate the process to make the deduced gravity and the assumed
gravity consistent.

We first turn to the deduced luminosities to determine how well they
scale from one model with the same shape to another. For each model
we divide the deduced luminosity at each inclination by the actual
luminosity. The results are presented in Figure \ref{fig1} for models
with the ratio of the polar to equatorial radius of 0.82, the most
rapidly rotating case for which the temperatures of all five masses
fall within the range allowed. We see that the curves all have the
same shape, but that the variation from pole to equator increases
slightly as the mass increases, particularly at small inclination.
While not perfect, the results in Figure \ref{fig1} are sufficient
to indicate that a reasonable determination of the intrinsic luminosity
could be made given an observed luminosity, inclination, and polar
to equatorial radius ratio (assuming uniform rotation), at least in
the mass range covered.

For the deduced effective temperatures, we proceed in a manner similar
to that used for the luminosity. Here the actual effective temperature,
defined as the effective temperature obtained from the flux given
by the actual luminosity divided by the total surface area of the
model, plays the role that the actual luminosity played in the previous
discussion. We take the ratio of the deduced effective temperature
at each inclination divided by the actual effective temperature for
each mass at a given shape. The results are shown in Figure \ref{fig2}.
Again we see that the curves for different masses show the same form.
Interestingly, the largest differences are shown for models seen equator
on instead of pole on, except for the $3M_{\odot}$ model, whose ratio
at low inclination is noticeably larger than that for all the other
masses. This variation with mass for both the deduced effective temperature
and deduced luminosity suggests that these might be analogous to homology
transforms for realistic models of stars \textendash{} it works well
over a restricted mass range, but is not universal and progressively
degrades as the physical properties of the models become less similar.
It is worth mentioning that the ratio of the model effective temperatures
at a given latitude for these masses does not show any significant
latitudinal variation, so that the variation in the deduced effective
temperatures must originate in the conversion from physical effective
temperatures to observed ones.

Finally we consider whether the bolometric corrections deduced from
these simulated SEDs are affected by any substantial changes rotation
may introduce into the SED. This would be important if a full SED
was not available. We computed the visual magnitude of our models
and calculated $M_{V}$ using the assumed distance. The visual magnitude
was calibrated by scaling the flux of our spherical model for Vega
to match the observed value above the earth\textquoteright{}s atmosphere
at $5556\textrm{\AA}$ (\citealt{Hayes1975}) and then integrating
the flux in the V filter and requiring $V=0.03$ mag (\citealt{Bessell1998}).
The absolute bolometric magnitude comes directly from the model luminosity
with the bolometric magnitude of the sun set to 4.74. The bolometric
corrections have been computed at inclinations between 0 and 90 degrees
in ten degree intervals for all models. The results are shown in Figure
\ref{fig3}, a plot of the bolometric corrections as a function of
(B-V) for all inclinations of all models at all rotation speeds. Also
shown in Figure 3 are the bolometric corrections for spherically symmetric
models, the stars for $\mathrm{log}\, g=4.333$ and the triangles
for $\mathrm{log}\, g=4.0$. Figure 3 indicates that the temperature
is the key to the determination of the bolometric correction, but
also that the effective gravity also plays a role (to about 0.07 magnitudes).
Except through the effective gravity, rotation by itself does not
appear to produce any particular modifications to the bolometric corrections
for these models.

\section{APPLICATION TO MODELS NOT ON THE ZAMS}

\begin{deluxetable}{ccccc}

\tabletypesize{\footnotesize}
\tablecolumns{5}
\tablewidth{0pt}
\tablecaption{ Model properties \label{tab:zmodels}}

\tablehead{
\colhead{Model}  & \colhead{Mass} & \colhead{$V_{\mathrm{eq}}$} & \colhead{Actual $T_{\mathrm{eff}}$} & \colhead{Actual $L$}
\\
\colhead{} & \colhead{(M$_{\odot}$)} & \colhead{(km/s)} & \colhead{(K)} & \colhead{(K)}
}

\startdata
ZAMS 1 & $2.25$ & $287$ & $9474.5$ & $25.48$
\\
ZAMS 2 & $1.85$ & $237$ & $8187.9$ & $11.71$
\\
$\alpha$ Oph & $2.19$ & $229$ & $8122.8$ & $32.63$
\enddata

\end{deluxetable}

The scaling relationship for deduced luminosities and deduced $T_{\mathrm{eff}}$
described above for the case of ZAMS models can be extended to models
that are not in the same evolutionary state. To demonstrate this we
consider a model of $\alpha$ Ophiuchus, a rapidly rotating A-type
star. Interferometric observations of $\alpha$ Oph imply that it
has a polar to equatorial radius ratio of 0.836 (\citealt{Monnier2010a})
and Vsin i in the range of 210 - 240 km $\mathrm{\mathrm{s}}{}^{-1}$
(e.g., \citealt{BernaccaP.L.1970,Abt1995,Royer2002}). We compared
this model with two ZAMS models which have the same shape as $\alpha$
Oph: one has a similar mass but different actual $T_{\mathrm{eff}}$
and the other with similar actual $T_{\mathrm{eff}}$ to $\alpha$
Oph but different mass. A summary of the properties of each model
is given in Table \ref{tab:zmodels}. Using as input the deduced luminosity
and effective temperature at a specific inclination from the $\alpha$
Oph model and applying the scaling relations to each ZAMS model allows
the determination of the deduced effective temperatures and luminosities
at all inclinations, the effective temperature as a function of latitude,
and finally the actual effective temperature and luminosity for the
$\alpha$ Oph model. We can then compare the results predicted by
the two ZAMS models with those for the $\alpha$ Oph model itself.

The comparisons of the deduced effective temperatures and actual effective
temperatures are shown in Figure \ref{fig4}. We show the deduced
temperatures as functions of inclination determined from the $\alpha$
Oph model (crosses), from the $2.25M_{\odot}$ model (circles), and
from the $1.85M_{\odot}$ model (squares). Somewhat in keeping with
the interferometric results for $\alpha$ Oph, we have chosen the
deduced effective temperature and luminosity to be at an inclination
of $90^{\circ}$. The latitudinal variation of the effective temperatures
of the model is represented by the dotted line, and the variation
of the deduced effective temperatues with inclination is shown by
the dashed line. The same quantities are presented for the scaled
$2.25M_{\odot}$ and $1.85M_{\odot}$ models. We note that in both
cases each of the scaled two ZAMS models agree well with the actual
model for $\alpha$ Oph. \textbf{These results suggest that the precise
details of the comparison model are not too important as long as the
interior structures are sufficiently homologous. While ``sufficiently
homologous'' is somewhat loosely defined, clearly these two models
fit the requirements. On the other hand, one would not expect a $10M_{\odot}$
main sequence star to be an appropriate model for either $\alpha$
Oph or for a $10M_{\odot}$ red giant.}

The result is the same for the deduced luminosity, as shown in Figure
\ref{fig5}. Both ZAMS models scale well to the deduced luminosity
for the $\alpha$ Oph model at all inclinations, although the $1.85M_{\odot}$
ZAMS model agrees with the $\alpha$ Oph model a little better. As
one might expect from the agreement of these features, the actual
luminosity and the actual effective temperature for each ZAMS model
agree to within $0.12\, L_{\odot}$ and 40K with the value of the
$\alpha$ Oph model.

\section{\label{sec:SCALING-ALGORITHM}SCALING ALGORITHM}

We have argued that the deduced luminosity and effective temperature
scale for rotating models with the same shape over some limited range
of conditions, and that this allows us to determine reasonable values
for the luminosity and actual effective temperature of the unknown
star from the models. Here we develop an algorithm to use the deduced
and model scaling relations to obtain some intrinsic properties of
a rotating star (for convenience, we shall refer to the rotating stellar
models with known properties as the \textquotedblleft{}model\textquotedblright{}
and the unknown object whose properties we wish to obtain as the \textquotedblleft{}star\textquotedblright{})
. We start by assuming that we have a deduced effective temperature,
deduced luminosity, and a measurement of Vsin$i$. For the moment
we assume that we also have the shape and inclination for the star
as well. The reference model must have the surface radius, effective
temperature, surface rotational velocity, and surface effective gravity
as functions of latitude, the deduced effective temperatures and luminosities
as functions of inclination, and the mass and luminosity of the model.

The algorithm proceeds as follows: from Vsin$i$ and the inclination
of the star, compute the surface equatorial velocity, $V_{eq}$. Compute
the ratio of the surface equatorial velocity of the model and the
star. Because equation (\ref{eq:1}) is true for the surface velocity,
one can compute the star's surface rotation velocity at all latitudes.
Because both the deduced effective temperatures and the latitudinal
effective temperatures scale as shown in Figure \ref{fig4}, we have
\begin{equation}
\frac{T_{\mathrm{eff},1}(\theta_{j})}{T_{\mathrm{eff},2}(\theta_{j})}=\frac{T_{\mathrm{d},1}(i_{j})}{T_{\mathrm{d},2}(i_{j})}=\frac{T_{\mathrm{eff,}a,1}}{T_{\mathrm{eff,}a,2}}\quad\forall\quad j
\end{equation}
here 1 refers to the model, 2 refers to the star, $d$ refers to the
deduced temperature, and $a$ refers to the actual $T_{\mathrm{eff}}$
previously defined. Because the deduced temperatures are known at
the inclination of the star for both the model (by interpolation)
and the star (by observation), we can obtain the effective temperature
at all latitudes and the actual effective temperature for the star.
The actual luminosity of the star can be computed from the deduced
luminosity at the assumed inclination, the deduced luminosity as a
function of inclination for the model and the actual luminosity for
the model. With the actual luminosity and actual effective temperatures
known for both the star and the model, one can compute the ratio of
the radii because the only difference between the surface areas of
the model and the star is the difference in the radius. Hence, the
radius of the star at every latitude of the star follows from the
radius profile of the model.

The steps so far have depended only on the model, its deduced properties,
and the scaling relations for both. These steps have resulted in the
surface properties of the star as a function of latitude. The next
step requires that the surface be an equipotential. As shown in Section
\ref{sub:Scaling-of-Model-properties}, we use the fact that the gravitational
part of the total potential and centrifugal potential must scale the
same way for a given shape (i.e., at a given latitude, the ratio between
the centrifugal potentials of the model and star and the ratio between
the gravitational potentials of the model and the star must be the
same). Because we have computed both the surface rotational velocity
and the surface radius as functions of latitude for the star and have
them for the model, we can obtain the ratio of the centrifugal potentials
and hence of the\textcolor{red}{{} }gravitational potentials. \textbf{Figure
6 presents a diagram of the information required from both the model
and the star, as well as how the various unknowns of the star are
determined from the scaling.}

If we assume that the gravitational potential at the surface is given
by that of a spherical star, at least at the equator, we can compute
an estimate for the mass of the star. This assumption is generally
good unless there is significant differential rotation where the material
close to the rotation axis rotates much faster than the material farther
away from the axis. It is certainly excellent for uniformly rotating
models except those very near critical rotation (e.g., \citealt{Ostriker1968,Faulkner1968,Jackson2004,Deupree2011a}).

The reason for performing these last steps was to get an estimate
of the mass which could be used with the actual luminosity to provide
a check on the results. All the models utilized here are core hydrogen
burning objects for which the main sequence mass - luminosity law
should hold. A check on the reasonableness of the assumed inclination
and shape can be made through how well the derived mass and actual
luminosity fit the mass-luminosity law.

\section{LIMITATIONS OF THE ALGORITHM}

The scaling we have described relies on certain assumptions, and it
is reasonable to see to what extent they can be relaxed. We have assumed
that the surface is an equipotential, which only exists if the rotation
law is conservative. Even for conservative rotation laws, it remains
an assumption that the surface is an equipotential. This likely matters
for the part of the solution that makes an estimate of the mass, but
it need not affect the scaling of the observable properties as long
as whatever mechanism determines the surface shape determines it in
the same way for both the unknown and comparison objects. Our very
limited knowledge of the surfaces of rotating stars does not allow
an answer to this question.

We also assume that we know both the inclination and the surface shape
of the unknown object. This in general is not true, and it turns out
that there are combinations of inclination and shape which produce
reasonable results, including masses which fit the mass-luminosity
relation. The general trend is that more rapid rotation (i.e., more
oblate shapes) can be offset by smaller inclinations. One might also
add that determining the surface shape accurately potentially pays
dividends by possibly placing constraints of the rotation law. 

We have also used a single composition for all our models. A different
composition would make a difference by producing a different color
- effective temperature relation. We can obtain a crude estimate of
how this might affect results by comparing the color-effective temperature
relations of several spherical models at different temperatures with
two different compositions. We calculated NLTE plane parallel model
atmospheres with temperatures of 8000, 9000, and 10000K using half
the metallicity of our previously computed models. The deduced temperatures
for spherical models at these temperatures were all within 50K of
the actual temperature when using the color-effective temperature
relation from the full metallicity models.

These results lead us to believe that scaling of observables can be
a useful technique to make the bridge between what one observes for
rotating stars and physically useful information under appropriate
conditions. We should caution that these results cover only a limited
range in gravity and effective temperature and that extension far
outside this range may not be warranted.

The authors thank Compute Canada and ACEnet for the computational
resources used in this research.

\bibliographystyle{apj}
\bibliography{bibliography}

\begin{thebibliography}{}
\expandafter\ifx\csname natexlab\endcsname\relax\def\natexlab#1{#1}\fi

\bibitem[{Abt \& Morrell(1995)}]{Abt1995}
Abt, H.~A., \& Morrell, N.~I. 1995, ApJS, 99, 135

\bibitem[{Aufdenberg {et~al.}(2006)Aufdenberg, Merand, du~Foresto, Absil, {Di
  Folco}, Kervella, Ridgway, Berger, ten Brummelaar, McAlister, Sturmann,
  Sturmann, \& Turner}]{Aufdenberg2006}
Aufdenberg, J.~P., Merand, A., du~Foresto, V.~C., {et~al.} 2006, ApJ, 645, 664

\bibitem[{Bernacca \& Perinotto(1970)}]{BernaccaP.L.1970}
Bernacca, P.~L., \& Perinotto, M. 1970, Contributions dell Osservatorio
  Astrofisica dell Universita di Padova in Asiago

\bibitem[{Bessell {et~al.}(1998)Bessell, Castelli, \& Plez}]{Bessell1998}
Bessell, M.~S., Castelli, F., \& Plez, B. 1998, A\&A

\bibitem[{Castelli \& Kurucz(1994)}]{Castelli1994}
Castelli, F., \& Kurucz, R.~L. 1994, A\&A, 281, 817

\bibitem[{Che {et~al.}(2011)Che, Monnier, Zhao, Pedretti, Thureau, M\'{e}rand,
  ten Brummelaar, McAlister, Ridgway, Turner, Sturmann, \& Sturmann}]{Che2011}
Che, X., Monnier, J.~D., Zhao, M., {et~al.} 2011, ApJ, 732, 68

\bibitem[{Claret(2012)}]{Claret2012}
Claret, A. 2012, Astron. Astrophys., 541, A113

\bibitem[{Collins \& Harrington(1966)}]{Collins1966}
Collins, George~W., I., \& Harrington, J.~P. 1966, ApJ, 146, 152

\bibitem[{Dall \& Sbordone(2011)}]{Dall2011}
Dall, T.~H., \& Sbordone, L. 2011, Journal of Physics Conference Series, 328,
  012016

\bibitem[{Deupree(1990)}]{Deupree1990}
Deupree, R.~G. 1990, ApJ, 357, 175

\bibitem[{Deupree(1995)}]{Deupree1995}
---. 1995, ApJ, 439, 357

\bibitem[{Deupree(2011)}]{Deupree2011a}
---. 2011, ApJ, 735, 69

\bibitem[{Deupree {et~al.}(2012)Deupree, Casta\~{n}eda, Pe\~{n}a, \&
  Short}]{Deupree2012}
Deupree, R.~G., Casta\~{n}eda, D., Pe\~{n}a, F., \& Short, C.~I. 2012, ApJ,
  753, 20

\bibitem[{{Domiciano de Souza} {et~al.}(2003){Domiciano de Souza}, Kervella,
  Jankov, Abe, Vakili, di~Folco, \& Paresce}]{DomicianodeSouza2003}
{Domiciano de Souza}, A., Kervella, P., Jankov, S., {et~al.} 2003, A\&A, 407,
  L47

\bibitem[{{Espinosa Lara} \& {Rieutord}(2011)}]{2011A&A...533A..43E}
{Espinosa Lara}, F., \& {Rieutord}, M. 2011, \aap, 533, A43

\bibitem[{Faulkner {et~al.}(1968)Faulkner, Roxburgh, \&
  Strittmatter}]{Faulkner1968}
Faulkner, J., Roxburgh, I.~W., \& Strittmatter, P.~A. 1968, ApJ, 151, 203

\bibitem[{Fr\'{e}mat {et~al.}(2005)Fr\'{e}mat, Zorec, Hubert, \&
  Floquet}]{Fremat2005a}
Fr\'{e}mat, Y., Zorec, J., Hubert, A.-M., \& Floquet, M. 2005, A\&A, 440, 305

\bibitem[{Gillich {et~al.}(2008)Gillich, Deupree, Lovekin, Short, \&
  Toqu\'{e}}]{Gillich2008}
Gillich, A., Deupree, R.~G., Lovekin, C.~C., Short, C.~I., \& Toqu\'{e}, N.
  2008, ApJ, 683, 441

\bibitem[{Hardorp \& Strittmatter(1968)}]{Hardorp1968}
Hardorp, J., \& Strittmatter, P.~A. 1968, ApJ, 151, 1057

\bibitem[{Hauschildt \& Baron(1999)}]{HauschildtP.H.1999}
Hauschildt, P.~H., \& Baron, E. 1999, JCoAM, 109

\bibitem[{Hayes \& Latham(1975)}]{Hayes1975}
Hayes, D.~S., \& Latham, D.~W. 1975, ApJ, 197, 593

\bibitem[{Jackson {et~al.}(2004)Jackson, MacGregor, \& Skumanich}]{Jackson2004}
Jackson, S., MacGregor, K.~B., \& Skumanich, A. 2004, ApJ, 606, 1196

\bibitem[{Linnell \& Hubeny(1994)}]{Linnell1994}
Linnell, A.~P., \& Hubeny, I. 1994, ApJ, 434, 738

\bibitem[{Lovekin {et~al.}(2006)Lovekin, Deupree, \& Short}]{Lovekin2006a}
Lovekin, C.~C., Deupree, R.~G., \& Short, C.~I. 2006, ApJ, 643, 460

\bibitem[{Maeder \& Peytremann(1970)}]{Maeder1970}
Maeder, A., \& Peytremann, E. 1970, A\&A, 7, 120

\bibitem[{Mirouh {et~al.}(2013)Mirouh, Reese, Lara, Ballot, \&
  Rieutord}]{IAU:9177147}
Mirouh, G.~M., Reese, D.~R., Lara, F.~E., Ballot, J., \& Rieutord, M. 2013,
  Proceedings of the International Astronomical Union, 9, 455

\bibitem[{Monnier {et~al.}(2010)Monnier, Townsend, Che, Zhao, Kallinger,
  Matthews, \& Moffat}]{Monnier2010a}
Monnier, J.~D., Townsend, R. H.~D., Che, X., {et~al.} 2010, ApJ, 725, 1192

\bibitem[{Monnier {et~al.}(2007)Monnier, Zhao, Pedretti, Thureau, Ireland,
  Muirhead, Berger, Millan-Gabet, {Van Belle}, {Ten Brummelaar}, McAlister,
  Ridgway, Turner, Sturmann, Sturmann, \& Berger}]{Monnier2007}
Monnier, J.~D., Zhao, M., Pedretti, E., {et~al.} 2007, Science, 317, 342

\bibitem[{Ostriker \& Mark(1968)}]{Ostriker1968}
Ostriker, J.~P., \& Mark, J. W.-K. 1968, ApJ, 151, 1075

\bibitem[{Royer {et~al.}(2002)Royer, Grenier, Baylac, Gomez, \&
  Zorec}]{Royer2002}
Royer, F., Grenier, S., Baylac, M.-O., Gomez, A.~E., \& Zorec, J. 2002, A\&A,
  393, 897

\bibitem[{Slettebak {et~al.}(1980)Slettebak, Kuzma, \& Collins}]{Slettebak1980}
Slettebak, A., Kuzma, T.~J., \& Collins, G.~W. 1980, ApJ, 242, 171

\bibitem[{van Belle {et~al.}(2001)van Belle, Ciardi, Thompson, Akeson, \&
  Lada}]{VanBelle2001}
van Belle, G.~T., Ciardi, D.~R., Thompson, R.~R., Akeson, R.~L., \& Lada, E.~A.
  2001, ApJ, 559, 1155

\bibitem[{Yoon {et~al.}(2008)Yoon, Peterson, Zagarello, Armstrong, \&
  Pauls}]{Yoon2008}
Yoon, J., Peterson, D.~M., Zagarello, R.~J., Armstrong, J.~T., \& Pauls, T.
  2008, ApJ, 681, 570

\bibitem[{Zhao {et~al.}(2009)Zhao, Monnier, Pedretti, Thureau, M\'{e}rand, {Ten
  Brummelaar}, McAlister, Ridgway, Turner, Sturmann, Sturmann, Goldfinger, \&
  Farrington}]{Zhao2009}
Zhao, M., Monnier, J.~D., Pedretti, E., {et~al.} 2009, ApJ, 701, 209

\end{thebibliography}

\clearpage{}

\figcaption{\label{fig1}Ratio of the deduced luminosity to the luminosity as
a function of inclination for ZAMS uniformly rotating models with
masses of $1.875M_{\odot}$ (circles), $2M_{\odot}$ (crosses), $2.25M_{\odot}$
(stars), $2.5M_{\odot}$ (triangles) and $3M_{\odot}$ (squares).
The ratio of the polar radius to equatorial radius is 0.82, corresponding
to a surface equatorial velocity of about 300 km s$^{-1}$. }

\figcaption{\label{fig2}Ratio of the deduced effective temperature to the actual
effective temperature (defined as the luminosity divided by the total
surface area) as a function of inclination for the same models as
in \ref{fig1}. Note the general agreement, although the results for
the $3M_{\odot}$ suggest that there are limits to the applicability
of the scaling.}

\figcaption{\label{fig3}The dots show bolometric corrections for the SEDs of
all models and inclinations. The stars denote plane parallel model
atmospheres with $\mathrm{log}\, g=4.333$ and the triangles denote
those with 4.000. Note that the bolometric corrections are not significantly
different between the rotating models and the spherical models.}

\figcaption{\label{fig4}Plot of the effective temperature as a function of latitude
(symbols on the dotted line) and the deduced effective temperature
as a function of inclination (symbols on the dashed line). The diamonds,
triangles, and crosses for the effective temperature refer to the
results scaled from the $2.25M_{\odot}$ ZAMS model, the results scaled
from the $1.85M_{\odot}$ ZAMS model, and the actual effective temperatures
for the evolved model. The circles, squares, and crosses for the deduced
effective temperatures refer to the results scaled from the $2.25M_{\odot}$
ZAMS model, the results scaled from the $1.85M_{\odot}$ ZAMS model,
and the actual effective temperatures for the evolved model. A temperature
range of $100K$ is also indicated and it is clear that all three
temperatures agree with each other for all cases considered.}

\figcaption{\label{fig5}The deduced luminosity as a function of inclination
for the evolved model based on the scaled $2.25M_{\odot}$ ZAMS model
(circles), the scaled $1.85M_{\odot}$ ZAMS model (squares), and for
the actual evolved model itself (crosses).}

\figcaption{\label{fig6}Diagram showing representation of the scaling algorithm
(see section \ref{sec:SCALING-ALGORITHM}). It shows the information
required from both the model and the star, as well as how the various
unknowns of the star are determined from the scaling are presented. }

\setcounter{figure}{0}

\clearpage{}
\begin{figure}
\includegraphics[scale=0.35]{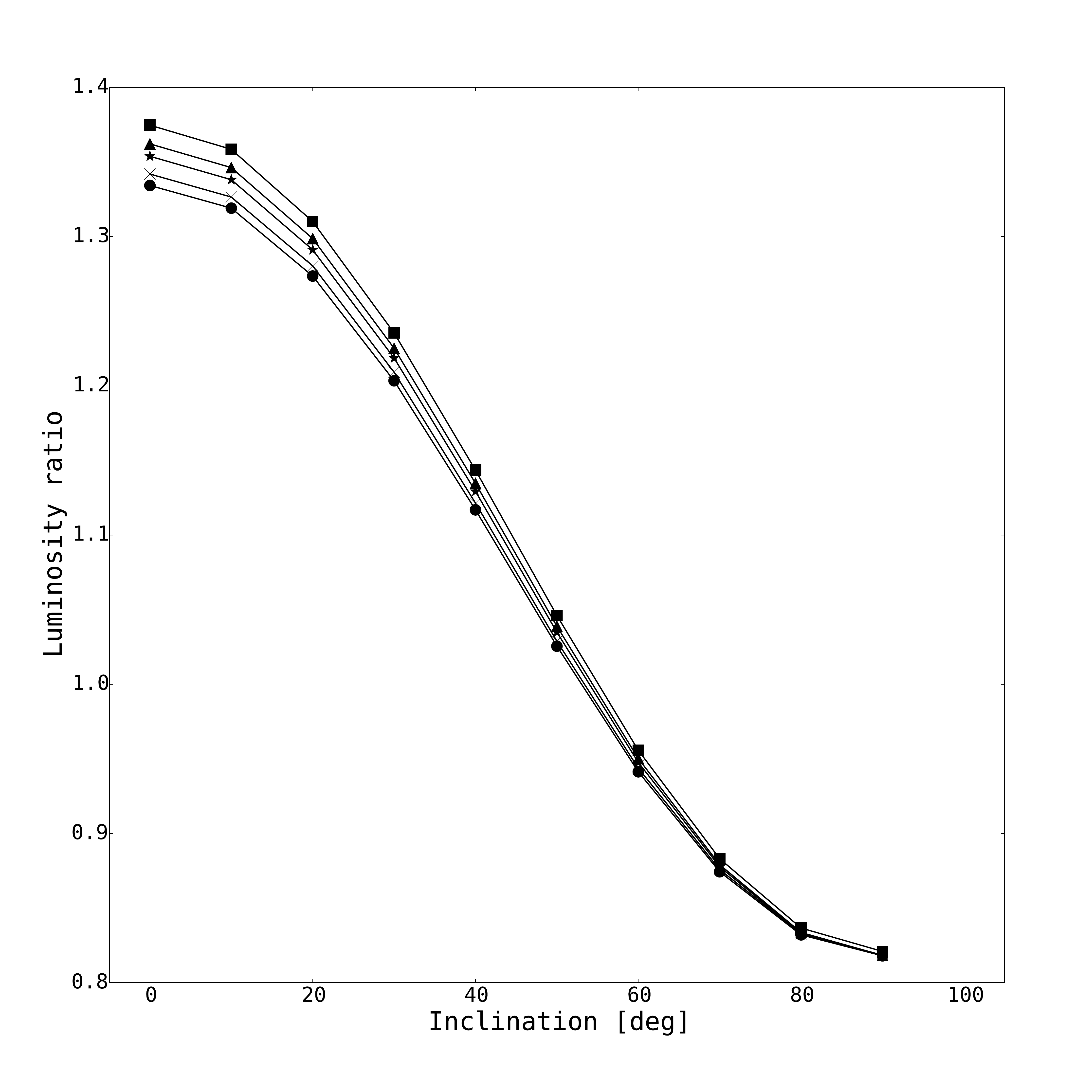}

\caption{}
\end{figure}

\clearpage{}
\begin{figure}
\includegraphics[scale=0.35]{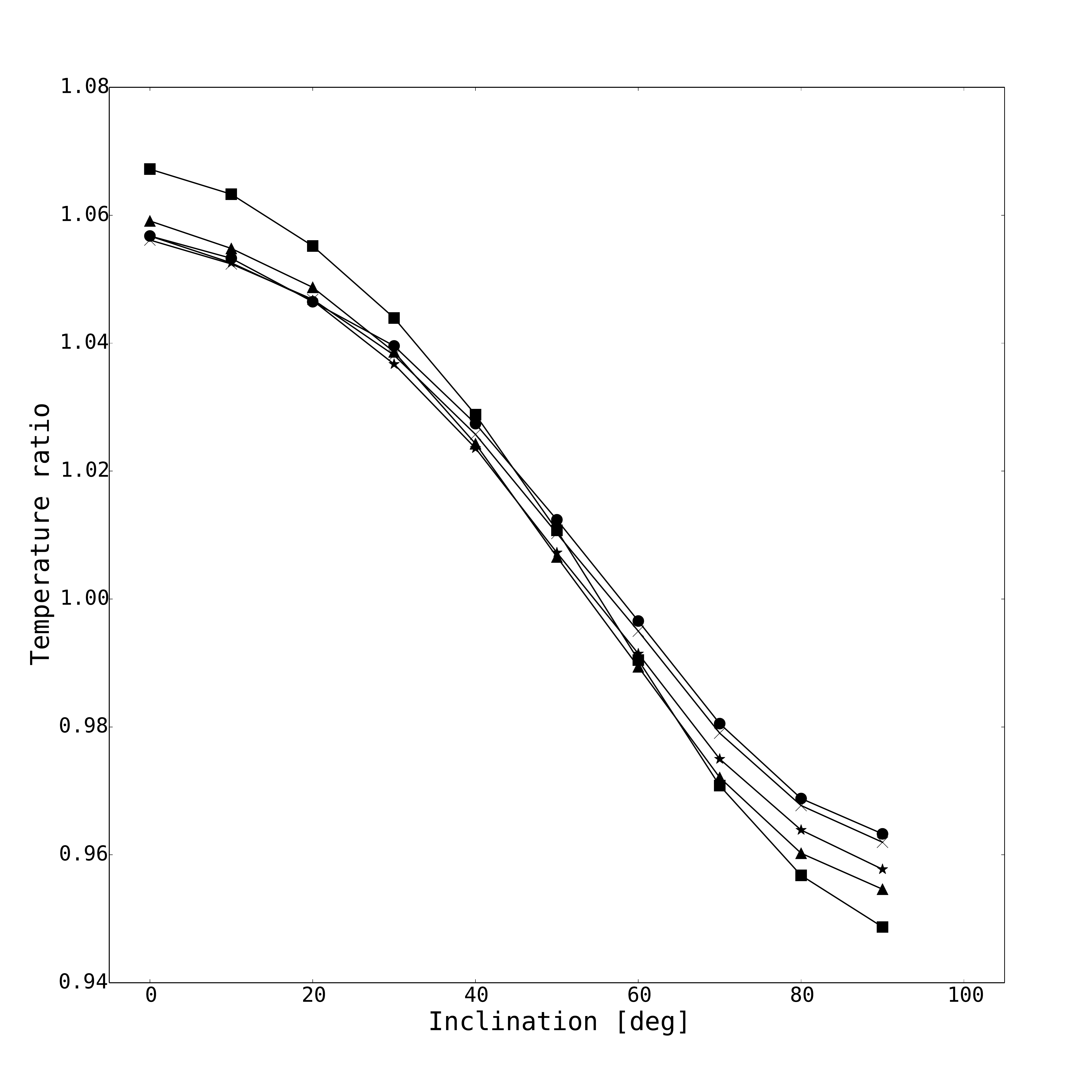}

\caption{}

\end{figure}

\clearpage{}
\begin{figure}
\includegraphics[scale=0.35]{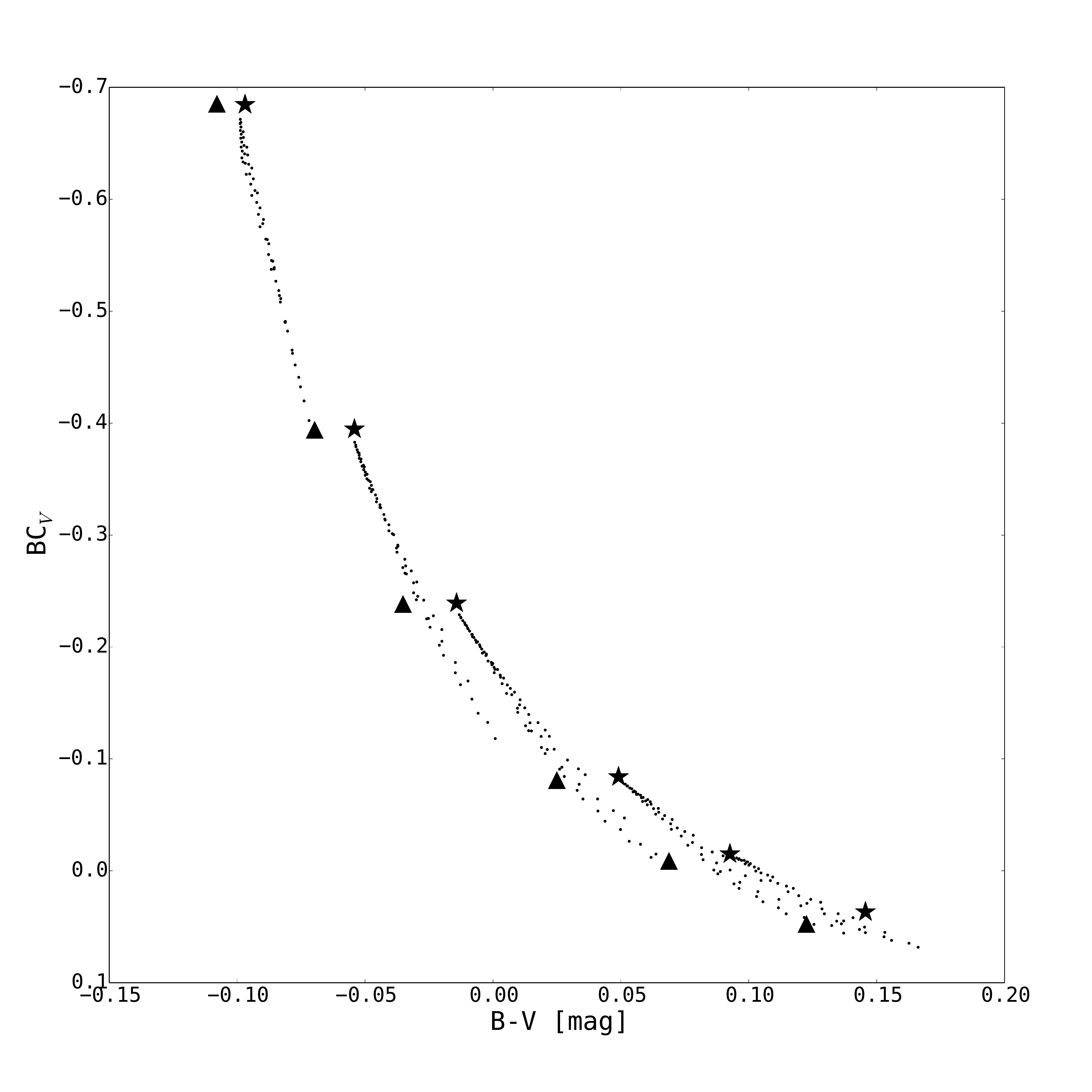}

\caption{}

\end{figure}

\clearpage{}
\begin{figure}
\includegraphics[scale=0.9]{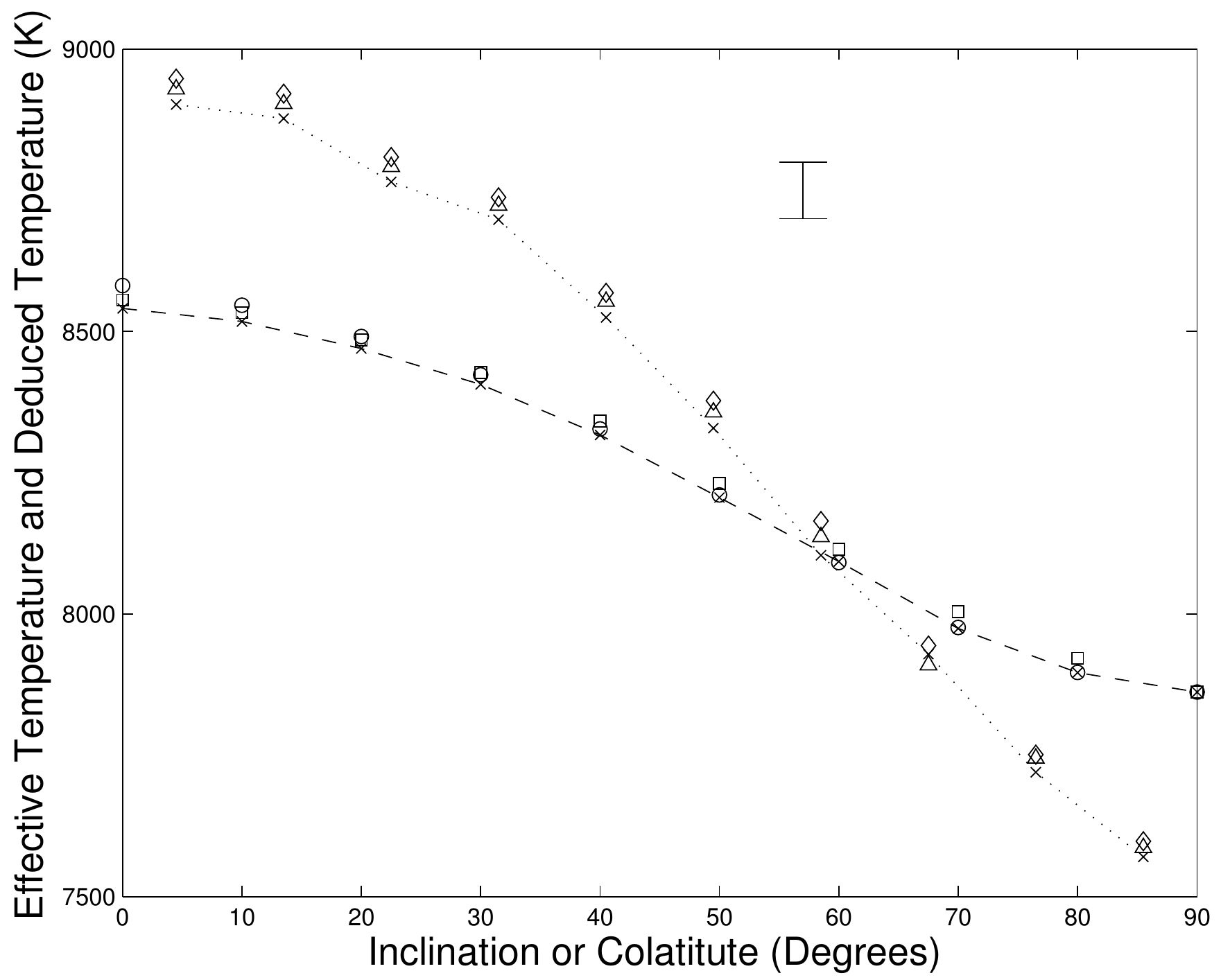}

\caption{}

\end{figure}

\clearpage{}
\begin{figure}
\includegraphics[scale=0.9]{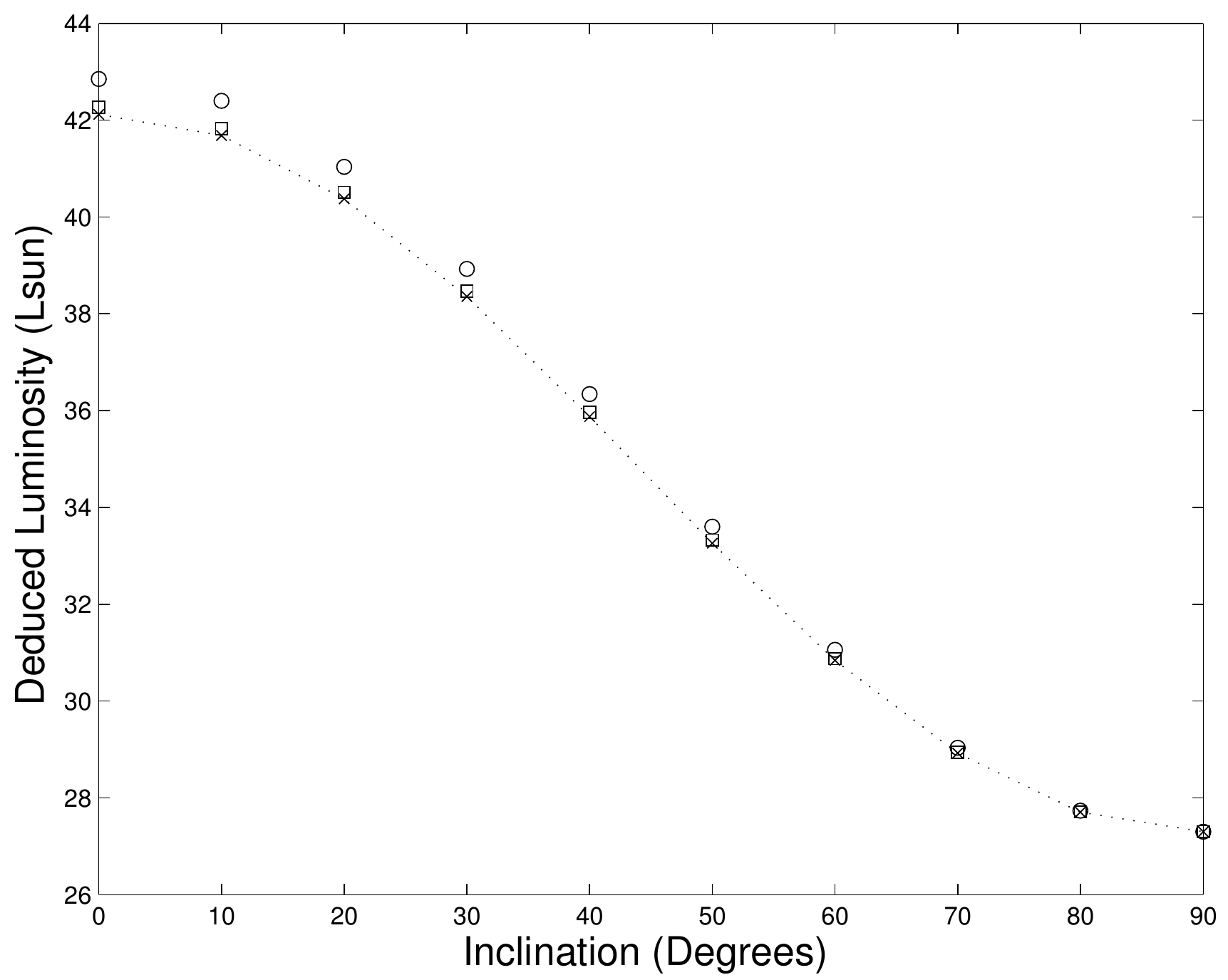}

\caption{}

\end{figure}

\clearpage{}
\begin{figure}
\includegraphics[scale=0.8]{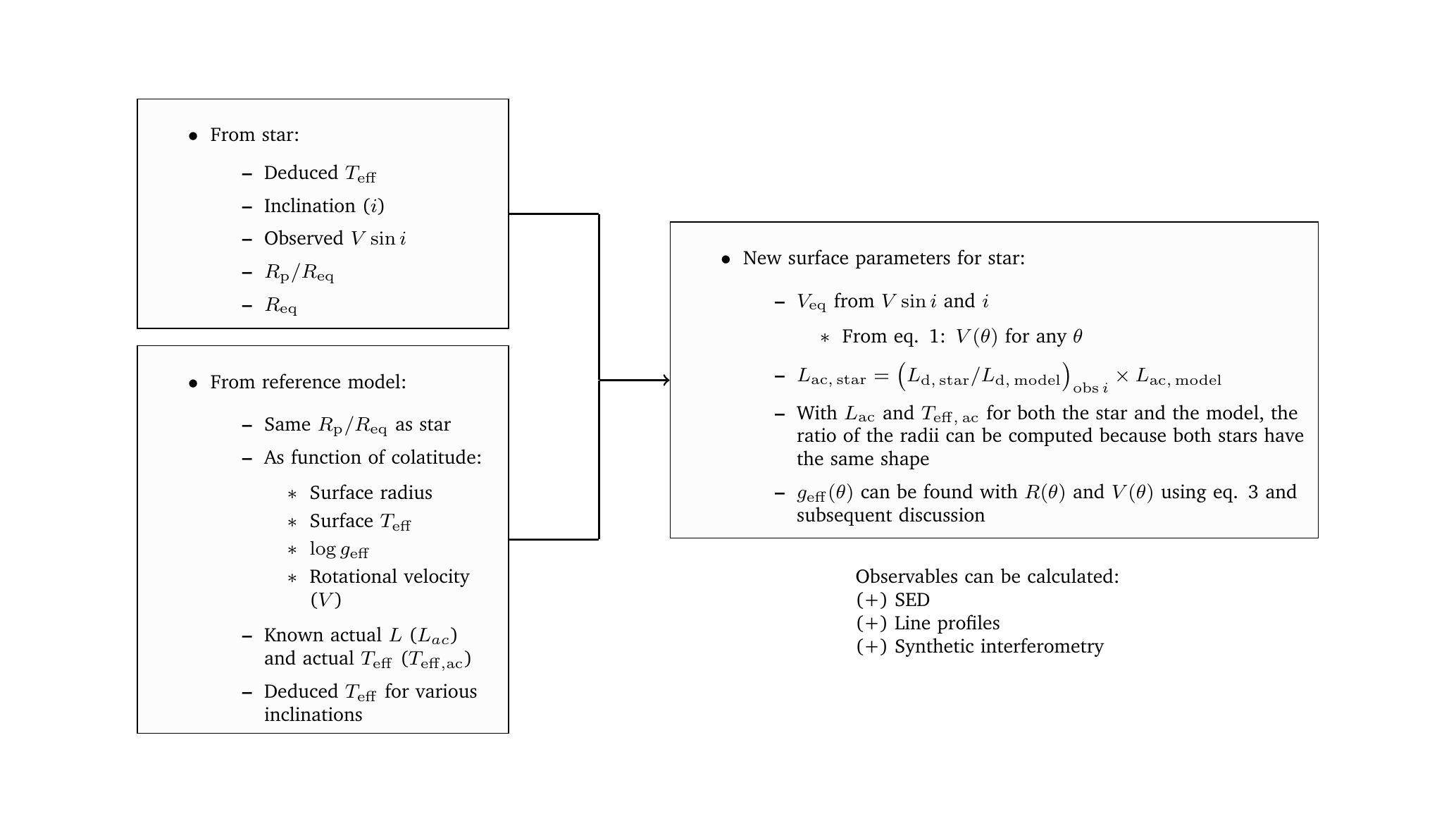}

\caption{}
\end{figure}

\end{document}